\documentclass[aps,twocolumn,showpacs]{revtex4}

\usepackage{graphicx}
\begin{document}
                         
\title{Charge Density Wave Driven Ferromagnetism 
in the Periodic Anderson Model}

\author{ M.~A. Majidi$^{1,2}$, D.~G.~S.~P. Doluweera$^{1}$,  B. Moritz$^{1,2}$,
 P.~R.~C. Kent$^{3}$, J. Moreno$^{2}$, and  M. Jarrell$^{1}$}
\affiliation{$^{1}$Department of Physics, University of Cincinnati, 
Cincinnati, Ohio 45221}
\affiliation{$^{2}$Department of Physics, University of North Dakota,
Grand Forks, ND 58202}
\affiliation{$^{3}$Center for Nanophase Materials Sciences, 
Oak Ridge National Laboratory, Oak Ridge, Tennessee 37831.}
\date{\today}

\begin{abstract}  
  We demonstrate the existence of ferromagnetism 
  in the Periodic Anderson Model (PAM) 
  at conduction-band filling near a quarter. 
  We show that this ferromagnetism is not supported 
  by Ruderman-Kittel-Kasuya-Yosida (RKKY) interactions but
  is instead driven by the precursors 
  of charge density wave (CDW) formation in the conduction electron band.
  To study the effect of spatial correlations, we compare
  Dynamical Mean field Approximation (DMFA) 
  and Dynamical Cluster Approximation (DCA) results.
  We find that both RKKY and CDW driven ferromagnetism persist as 
  short-range correlations are incorporated into the theory. 
  Both DMFA and DCA show the precursors of 
  CDW formation through the strong enhancement of the $d$-electron 
  CDW susceptibility as the temperature decreases, up to the ferromagnetic 
  transition temperature. In addition, the DCA captures the signal of 
  a band gap opening due to Peierls instability.
\end{abstract}


\maketitle

Rare-earth and actinide-based heavy fermion (HF) systems display a 
wide variety of interesting phenomena, 
including Kondo insulator, fermi and non-fermi liquid paramagnetism,
ferromagnetism, antiferromagnetism and superconductivity \cite{HFBooks}.
Among these phenomena, ferromagnetism is especially interesting since
while a number of HF ferromagnetic compounds exist 
\cite{HF-FM-examples}, few experimental
studies on the ferromagnetic phase diagrams are available 
\cite{Larrea_PRB05}. Although a number of theoretical studies of HF 
ferromagnetism are available 
\cite{Sigrist_PRB92,Sigrist_PhysB92,Troyer_PRB93,Moller_PRB93,
Doradzinski_PRB98,Meyer_PSS98,Meyer_EPJB00,Batista_PRB01,
Batista_PRB03,Gulacsi_PRB05},  
the range of ferromagnetism in the local moment regime 
has not been fully clarified, and the mechanisms leading to  
ferromagnetic order are still not well understood.

In this Letter, we investigate charge density wave (CDW) driven ferromagnetism 
occurring near a quantum critical point (QCP) around quarter filling of the 
conduction band in the Kondo regime of the Periodic Anderson Model (PAM).  
The ferromagnetism in the region of low conduction band filling (e.g. 
$n_d \ll 0.5$) is attributed to the Ruderman-Kittel-Kasuya-Yosida (RKKY)
exchange interaction between the $f$-electron local moments mediated by
the conduction ($d$-) electrons 
\cite{Sigrist_PRB92,Sigrist_PhysB92,Troyer_PRB93,Moller_PRB93,
  Doradzinski_PRB98,Meyer_PSS98,Meyer_EPJB00,Batista_PRB01,
Batista_PRB03,Niki_PRB97}. 
Whereas at $n_d\approx1$ the RKKY and superexchange interactions 
lead to antiferromagnetic order \cite{Niki_PRB97,JarrellPAM_PRB95}.
However, in the region near quarter filling $n_d\approx0.5$, close to a
QCP associated with a vanishing ferromagnetic transition temperature, 
the mechanism leading to magnetic order is not well understood. 
We provide theoretical evidence arising from the incorporation of 
spatial correlations into the theory that the ordering near quarter filling
of the conduction band in the PAM is 
enhanced by CDW fluctuations. Previous studies (e.g. \cite{Meyer_PSS98} and 
\cite{Batista_PRB01}) likely did not find ferromagnetic ordering 
at this filling due to the difficulty of capturing strong CDW correlations 
with earlier methods.

Our interest is enhanced by the discovery of the ferromagnetic 
superconductor UGe$_2$ \cite{Saxena_Nature00}, where superconductivity 
coexists with ferromagnetism near its ferromagnetic QCP.
Pressure-dependence studies indicate that there may be different
mechanisms driving the ferromagnetic order in this material 
\cite{Nakane_JPSJ04,Onuki_JPSJ03}.
One scenario suggests that the superconducting-ferromagnetic coexistence
is related to coupled charge and spin density waves 
\cite{Watanabe-Miyake_PhysB02}.
Although our model is not detailed enough to describe Uranium-based compounds,
our results suggest that a CDW mechanism might play an important role 
in driving the ferromagnetic order near the QCP in UGe$_2$.

\begin{figure}
\includegraphics[width=3.2in]{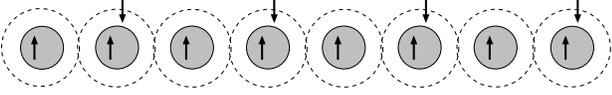}
\caption{Illustration of a $d$-electron  
charge density wave for $n_f=1$ and $n_d=0.5$ which mediates alignment
 of the $f$-electron local moments through coherent Kondo screening.
 Smaller shaded circles represent the $f$ orbitals, and bigger 
 dashed circles the $d$ orbitals. Up and down arrows represent 
 the electron spin orientations. }
\label{Cartoon}
\end{figure}

As briefly mentioned in a previous study of the infinite dimensional
PAM \cite{Niki_PRB97}, Kondo screening plays an important role in triggering 
the ferromagnetism when the RKKY and superexchange interactions favor 
antiferromagnetism.  We refer to this as charge density wave (CDW) driven 
ferromagnetism, since the alignment of the $f$-electron local moments is 
caused by coherent Kondo screening induced by the conduction electrons 
forming a CDW. Fig. \ref{Cartoon} illustrates this process.
Imagine there are roughly $N/2$ localized ($f$) electrons and
$N/4$ itinerant ($d$) electrons in the system, where $N$ is
the number of lattice sites. At some low temperature each 
$d$ electron locally screens one $f$ electron
through antiferromagnetic alignment of their spins. These local screening 
clouds \cite{Sorensen_PRB96,KondoCloud} become larger at lower 
temperatures. As a result, the $N/4$ $d$ electrons tend to screen all 
$N/2$ $f$ electrons coherently. 
This process requires that the $d$ electrons form 
a spin-polarized CDW, since in order to maximize kinetic energy gain and  
maintain the coherent Kondo screening a $d$ electron must hop to
nearest neighbor sites with  unoccupied $d$ orbitals.

{\em{Formalism.}}  Since Dynamical Mean Field Theory (DMFT) \cite{DMFTReview} 
in infinite dimensions has a local self energy, 
questions arise as to whether the CDW ferromagnetic 
mechanism persists in a finite dimensional system
and how spatial correlations may affect the results. 
In this Letter, we present our study on the three-dimensional PAM 
using the Dynamical Cluster Approximation (DCA) \cite{DCAReview}. 
We find that the CDW driven ferromagnetism persists as short-range spatial
correlations are incorporated into the theory.
In addition, we present one and two-particle evidence
of the precursors of CDW formation to support this ferromagnetic mechanism.

The PAM Hamiltonian can be written as
\begin{eqnarray}
\label{eq:Hamiltonian}
 H=-t\sum_{<ij>\sigma}(d^\dagger_{i \sigma}d_{j \sigma} + h.c.)
+ \sum_{i \sigma}(\epsilon_d d^\dagger_{i \sigma}d_{i \sigma} +
  \epsilon_f f^\dagger_{i \sigma}f_{i \sigma})
\nonumber \\
+ V\sum_{i\sigma}(d^\dagger_{i\sigma}f_{i \sigma} + h.c.)
+ \sum_i U(n_{fi\uparrow} - \frac{1}{2})
          (n_{fi\downarrow} - \frac{1}{2}), \nonumber \\
\end{eqnarray}
where $d(f)^{(\dagger)}_{i \sigma}$ destroys(creates) a $d(f)$ electron
at site $i$ with spin $\sigma$,
$t$ is the nearest-neighbor {\it d}-electron 
hopping, $\epsilon_d$ and $\epsilon_f$ the orbital energy of the {\it d} and
{\it f} electrons respectively, $V$ the {\it d-f} hybridization, 
and $U$ the on-site Coulomb repulsion between 
$f$ electrons. In this Hamiltonian we set the chemical potential to zero.

We work with a simple cubic lattice with bare conduction band-width 
$W=12t=3$ ($t=0.25$).  We choose $U=1.5W$ and $V=W/3$ to satisfy the 
Zlati\'{c}-Horvati\'{c} criterion  for the strong-coupling regime 
\cite{Zlatic-Horvatic_PRB83}: $u \equiv U/\pi\Gamma > 2.0$, 
where $\Gamma=V^2N(0)$ 
and $N(0)$ is the bare conduction band density of states at the Fermi level. 
The Hirsch-Fye Quantum Monte Carlo (QMC) algorithm \cite{Hirsch-Fye_PRL86} 
is used to solve the DCA cluster problem\cite{QMC-DCA_PRB01}. We perform 
calculations for two different cluster sizes: $N_c=1$, equivalent to DMFA, 
and $N_c=14$.
We choose the bipartite cluster $14A$, which is considered the first ``good'' 
cluster with $N_c>1$, following the selection criterion 
of Betts  \cite{Betts_CJP97,PKent_PRB05}.  Following the procedure in 
Ref.~\cite{Niki_PRB97}, for every $n_d$ value, we choose 
$\epsilon_d-\epsilon_f$ such that $n_f\approx1$ at a moderate temperature, 
say $\beta=1/T=10$, and keep the value of $\epsilon_d-\epsilon_f$ fixed as 
the temperature is decreased.   The values of $\epsilon_d$ and $\epsilon_f$ 
are adjusted (keeping $\epsilon_d-\epsilon_f$ fixed) during the 
self-consistent iterations to satisfy $n_f\approx1$ and the chosen value 
of $n_d$.  To study the systematic errors associated with the imaginary 
time increment ($\Delta \tau$) in the QMC we compare the results with
$t\Delta \tau=1/4$ and $t\Delta \tau=1/16$.

\begin{figure}
\includegraphics[width=3.2in]{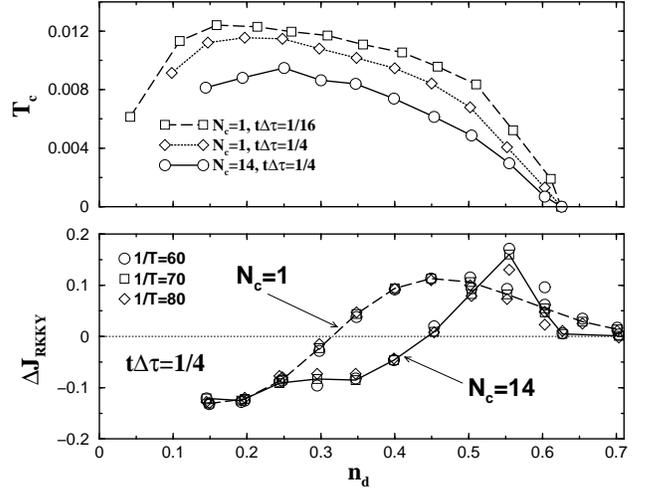}
\caption{Upper graph: Phase diagram showing $T_c$ as a function of $n_d$
 at $n_f=1$. The $T_c$ values are extracted from the $f$-electron 
 ferromagnetic susceptibility data.
 Lower graph: The corresponding $\Delta J_{RKKY}$ vs $n_d$. 
 The ferromagnetic phase diagram extends into the region where the 
 RKKY coupling is antiferromagnetic. The ferromagnetic region of 
 RKKY coupling extends toward higher $n_d$ values as $N_c$ is increased, 
 suggesting that the Fermi surface may be altered due to the introduction of
 ${\bf k}$-dependence in the self energy.}
\label{PhaseDiagram_and_DJRKKY}
\end{figure}

{\em{Results.}} The upper panel of Fig.~{\ref{PhaseDiagram_and_DJRKKY}} 
shows the FM Curie temperature, $T_c$, versus $n_d$, with $T_c$ vanishing 
at $n_d$ slightly higher than 0.6. Comparison between results 
for $N_c=1$ with  $t\Delta \tau=1/4$ and $t\Delta \tau=1/16$ suggests
that  $t\Delta \tau=1/4$ is sufficient to capture the correct profile 
of the phase diagram.  Comparison between results 
for $N_c=1$ and $N_c=14$ with the same $t\Delta \tau=1/4$ 
indicates that spatial correlations decrease $T_c$, 
as generally expected \cite{DCAReview}. 
The lower panel of Fig.~{\ref{PhaseDiagram_and_DJRKKY}} 
shows the dependence of $\Delta J_{RKKY}$ on $n_d$. 
Here $\Delta J_{RKKY}$ is defined as
$\Delta J_{RKKY}=J_{RKKY}^{AFM}-J_{RKKY}^{FM}$, 
with $J_{RKKY}^{AFM}=J_{RKKY}({\bf q}=(\pi,\pi,\pi))$ and 
$J_{RKKY}^{FM}=J_{RKKY}({\bf q}=(0,0,0))$, so that a negative (positive)
$\Delta J_{RKKY}$ signals a ferromagnetic (antiferromagnetic) RKKY coupling 
between neighboring local moments.
We define $J_{RKKY}({\bf q})\approx -2J_{fd}^2T\frac{1}{N}\sum_{{\bf k},n}
G^d({\bf k},i\omega_n) G^d({\bf k}+{\bf q},i\omega_n)$, 
with $J_{fd}\approx 8V^2/U$, as in Ref.~\cite{JarrellPAM_PRB95}.
For both $N_c=1$ and $N_c=14$, we take $t\Delta \tau=1/4$, and show that 
for three different temperatures close to
$T_c$ ($T$=1/60, 1/70, and 1/80) there is no significant difference in 
$\Delta J_{RKKY}$.  It is clear that for both values of $N_c$ the RKKY 
coupling changes from ferromagnetic to antiferromagnetic in a region 
where the phase is still ferromagnetic. In this region, 
where $\Delta J_{RKKY}>0$, ferromagnetism can not be explained by 
the RKKY mechanism.     
Note that the region of ferromagnetism supported by the RKKY 
mechanism becomes wider as $N_c$ changed from 1 to 14. Since, for a 
given $n_d$, RKKY coupling favors ferromagnetic alignment when
the magnitude of the Fermi vector satisfies $k_F < \pi/4$, this result 
suggests that the Fermi surface shrinks as spatial
correlations are introduced.

\begin{figure}
\includegraphics[width=3.2in]{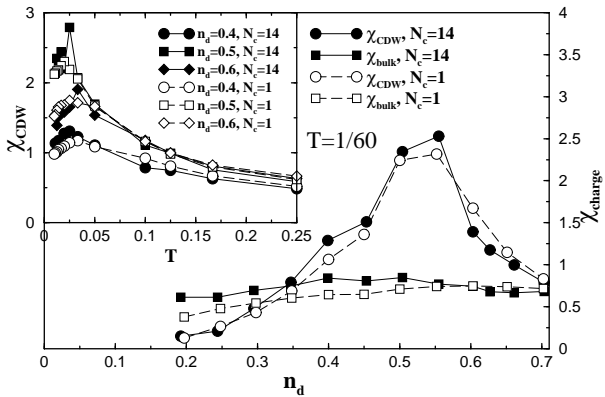}
\caption{
 The main panel shows the $d$-electron charge density wave (CDW)
 susceptibility and the bulk charge susceptibility as functions
 of $n_d$ at $T=1/60$, temperature at which the CDW susceptibility
 is strongly enhanced as indicated in the inset.
 The region of $n_d$ values where the CDW susceptibility is greater than
 the bulk charge susceptibility covers the non-RKKY region in 
 the ferromagnetic phase diagram.}
\label{Chi_CDW_and_Bulk}
\end{figure}

To address the mechanism of ferromagnetic order in the region of the phase
diagram where RKKY interactions favor antiferromagnetism, we provide evidence 
that in this region ferromagnetism is driven by the precursors of charge 
density wave formation. 
In support of this idea, Fig.~\ref{Chi_CDW_and_Bulk} shows that at low 
temperatures the $d$-electron CDW susceptibility, 
$\chi_{CDW}\equiv \chi_{charge}({\bf q}=(\pi,\pi,\pi))$, 
is strongly enhanced relative to the `bulk' (uniform) 
charge susceptibility,  
$\chi_{Bulk}\equiv \chi_{charge}({\bf q}=(0,0,0))$, 
for $n_d$ values corresponding to the 
ferromagnetic phase region  not supported by the RKKY mechanism. 
Based on geometry one expects the CDW configuration be best 
achieved when $n_d=0.5$, which implies that $\chi_{CDW}$ would be
most strongly enhanced for $n_d=0.5$. However, our results show that 
the maximum enhancement occurs around $n_d\approx 0.55$, 
which may, again, indicate shrinking of the Fermi surface.
In the inset of  Fig.~\ref{Chi_CDW_and_Bulk} we plot  
$\chi_{CDW}$ versus $T$ for $n_d=0.4$, 0.5, and 0.6, 
showing its  enhancement at low temperatures. 
In these calculations, $t\Delta \tau=1/4$.  
For smaller values of $t\Delta \tau$ (not shown),
$\chi_{CDW}$ increases, but the peak location does not shift.

To elaborate further on how the CDW driven ferromagnetism works,
note that, as shown in the inset of Fig.~\ref{Chi_CDW_and_Bulk},
$\chi_{CDW}$ reaches a maximum at temperatures around $T\approx 1/40-1/60$,
and decreases slightly when the temperature is lowered close to
$T_c$. The non-local correlations enhance $\chi_{CDW}$, and we also find
that they enhance the screened local moment for $n_d \approx 0.5$, while they
suppress it at other fillings (not shown).  This  may be interpreted as
an attempt to balance the kinetic and potential energies of the Kondo lattice.
The system can optimize the exchange energy by localizing a $d$ electron in a
singlet with the $f$ moment on half the sites.   This is balanced by the
kinetic energy in the usual way
by allowing the $d$ electrons to delocalize and screen other $f$-electron local
moments forming Kondo ``clouds''.   However, when $n_d \approx 0.5$, the
system may gain additional kinetic energy though the gap that accompanies
a $d$-electron charge density wave (CDW).  Note that this is not a
well-defined CDW  phase, since such an ordering would be manifest in the
divergence of $\chi_{CDW}$, and a full gap would suppress Kondo screening.
As the temperature is lowered further, the local Kondo screening clouds grow, 
so that the overlap between the Kondo clouds drive the $d$ electrons 
to collaborate to screen the $f$ electrons coherently. 
In doing so, each $d$ electron forms a Kondo singlet with an $f$ electron 
at a site only momentarily, then hops to one of its neighboring sites 
to regain the resonance with the $f$ electron at that site.
This process forces the $f$-electron local moments to align, hence the system 
forms a ferromagnetic phase. The momentary breaking of the Kondo singlet 
due to hopping of the $d$ electrons to their neighboring sites decreases 
their staggered charge correlation. Thus, this explains why $\chi_{CDW}$ 
is suppressed slightly as the temperature approaches $T_c$.

\begin{figure}
\includegraphics[width=3.2in]{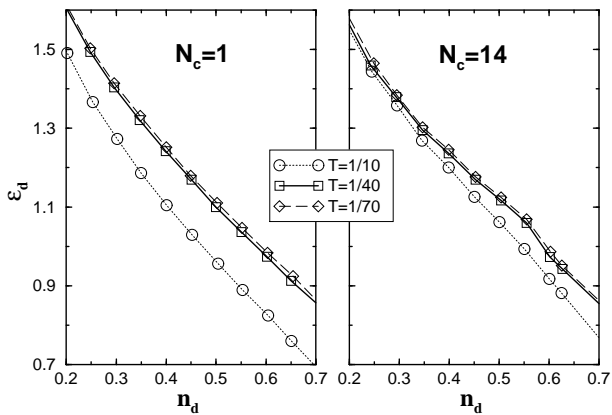}
\caption{Comparing $\epsilon_d$ versus $n_d$ for $N_c=14$ and $N_c=1$.
 For $N_c=14$ (right panel)  bending in the $\epsilon_d$ vs $n_d$
 curve is observed, indicating the tendency to the formation of a band 
 gap due to Peierls' instability. This feature is not seen for $N_c=1$
 (left panel).
 }
\label{ed_vs_nd}
\end{figure}

So far we have shown that DMFA can already capture the $d$-electron 
CDW precursors  at the {\it two-particle} level, i.e. 
in the $\chi_{CDW}$ behavior.
In the following, we show that by incorporating the spatial correlations 
within DCA ($N_c>1$), 
the CDW precursors can also be captured at the {\it one-particle} 
level, e.g. through the  $d$-electron density of states (DOS).
In order to capture the effective modulation potential
inducing the Peierls instability \cite{PeierlsBook}, 
the self energy must have sufficiently strong ${\bf k}$ dependence.
This is only possible for $N_c \gg 1$. 
When this modulation potential is properly captured by the self energy, 
the corresponding $d$-electron DOS forms a gap or 
signals of a gap opening, such as a ``dip", around the 
chemical potential. In our model, the chemical potential is set fixed 
to zero, but we adjust $\epsilon_d$ to satisfy the desired filling. 
Therefore, if a gap or a ``dip" occurs in the DOS, it should 
be manifest in a plot of  $\epsilon_d$ versus $n_d$. 
Fig.~\ref{ed_vs_nd} shows how  the $\epsilon_d$ vs $n_d$ profile evolves 
from high to low temperatures for $N_c=1$ and  $N_c=14$. It is clear that
for $N_c=14$ (right panel), but not for $N_c=1$ (left panel), 
the $\epsilon_d$ vs $n_d$ curve bends slightly 
near $n_d\approx$ 0.5 - 0.55 at $T\approx 1/40$,
which coincides with the strongest enhancement in 
$\chi_{CDW}$. We interpret this as the signal of a gap opening 
accompanying the formation of a CDW in
the $d$-electron DOS. 
As the temperature is lowered further, e.g. from
$T\approx 1/40$ to $T\approx 1/70$,
the bending in the $\epsilon_d$ vs $n_d$ curve does not
become more prominent, and diminishes 
as the temperature becomes closer to $T_c$, e.g. 
$T\approx 1/90$ (not shown). This indicates that
the gap never fully opens, and the $d$-electron DOS increases back 
as the $d$ electrons become more mobile to screen coherently at temperatures 
closer to  $T_c$.

Our arguments may be justified also by inspection of  
the Kondo screening length \cite{Sorensen_PRB96,KondoCloud} 
and its consistency with the 
$\chi_{CDW}$ and the $d$-electron density of
states as discussed before.
The magnitude of the Kondo coupling in our system is
$J_{fd}\approx 8V^2/U\approx 0.6W\approx 7t$.
According to the scaling theory
of Sorensen {\it et al.}  \cite{Sorensen_PRB96},
this coupling strength roughly corresponds to the
Kondo screening length, $\xi_L$, of the order of a lattice
constant or less. This is consistent with our picture of localized 
Kondo `clouds' when the CDW  forms.  
This screening length depends on the $d$-electron density of
states at the chemical potential, $\rho(0)$, through
$\xi_L \propto \exp[-1/\rho(0)J_{fd}]$. 
Since $\rho(0)$ increases as the temperature approaches $T_c$ 
(as discussed in the previous paragraph), the screening 
length also increases accordingly. In turn, the overlap between
screening clouds enables the $d$ electrons 
to mediate ferromagnetic coupling between
$f$-electron local moments.

{\em{Conclusion.}} We have shown that the ferromagnetism in the 
strong coupling regime of the PAM has two mechanisms. In the region of 
low conduction band filling ferromagnetism is driven by 
the RKKY exchange interaction, while in the region of higher conduction 
band fillings up to slightly more  than a quarter the precursors of 
a CDW formation in the $d$ electrons induce the magnetic order. 
We have demonstrated that the CDW formation can 
be captured within the DMFA and DCA through the enhancement of 
$d$-electron CDW susceptibility. In addition, the DCA captures the 
signal of a band gap opening due to the Peierls instability that 
drives the CDW formation.

We thank Th. Maier, A. Macridin, and K.\ Mikelsons for their
contributions to the codes used for our calculations. This research
was supported by NSF grants DMR-0312680, DMR-0706379,  and
DMR-0548011. A portion (PRCK) was conducted at the Center for Nanophase
Materials Sciences, which is sponsored at Oak Ridge National
Laboratory by the Division of Scientific User Facilities,
U.S. Department of Energy. Computation was carried out under
TeraGrid project TG-MCA06N019 and at the University of North
Dakota Computational Research Center, supported by EPS-0132289 and
EPS-0447679.

\end{document}